\documentclass[conference]{IEEEtran}

\IEEEoverridecommandlockouts
\usepackage{cite}
\usepackage{amsmath,amssymb,amsfonts}
\usepackage{algorithmic}
\usepackage{graphicx}
\usepackage{textcomp}
\usepackage{xcolor}
\usepackage{fancyhdr}
\usepackage{comment}
\usepackage{optidef}
\newtheorem{definition}{Definition}
\usepackage{graphics}
\usepackage{booktabs}
\usepackage{multirow}
\usepackage{graphicx}

\IEEEoverridecommandlockouts

\begin{document}

\title{Exploring Physical-Based Constraints in Short-Term Load Forecasting: A Defense Mechanism Against Cyberattack }

\author{\IEEEauthorblockN{Mojtaba Dezvarei, Kevin Tomsovic, Jinyuan Stella Sun, Seddik M. Djouadi
\thanks{This work was partially supported by NSF grant CNS-2038922. Mojtaba Dezvarei, Kevin Tomsovic, Jinyuan Stella Sun, and Seddik M. Djouadi
are with the Department of Electrical Engineering and Computer Science, The University of Tennessee, Knoxville, TN 37996 USA. (e-mail: mdezvare@utk.edu; tomsovic@utk.edu; jysun@utk.edu; mdjouadi@utk.edu)}
}}


\maketitle

\thispagestyle{fancy}

\IEEEpubidadjcol
\begin{abstract}
Short term load forecasting is an essential task that supports utilities to schedule generating sufficient power for balancing supply and demand, and can become an attractive target for cyber attacks. It has been shown that the power system state estimation is vulnerable to false data injection attacks. Similarly, false data injection on input variables can result in large forecast errors. The load forecasting system should have a protective mechanism to mitigate such attacks. One approach is to model  physical system constraints that would identify anomalies. This study investigates possible constraints associated with a load forecasting application. Looking at regional forecasted loads, we analyze the relation between each zone through similarity measures used in time series in order to identify constraints. Comprehensive results for historical ERCOT load data indicate variation in the measures recognizing the existence of malicious action. Still, these static measures can not be considered an efficient index across different scenarios.
\end{abstract}


\section{Introduction} 
Load forecasting is a critical to provide an efficient decision-making in power system operation and planning. Among the load forecasting types, short term load forecasting (STLF) plays a significant role in achieving economic, reliable, and secure operating operation. STLF  supports a time horizon ranging from one hour to one week. The primary objective of the STLF function is to provide load predictions for the basic generation scheduling functions (unit commitment and hydroscheduling), for assessing the security of the power system at anytime point, and timely dispatcher information \cite{gross1987short}. \par
To predict the electric load, there are critical variables that accurately forecast the load. Many forecasting models with a wide range of methodologies for STLF are suggested in the literature. The models are mainly classified into two groups: statistical techniques, such as, multiple linear regression models (MLR), autoregressive and moving average (ARMA) models; and artificial intelligence (AI) techniques, such as, artificial neural network (ANN), fuzzy regression models, support vector machines (SVMs) \cite{hong2016probabilistic}. All methods are constructed based on input data indicating the independent variables compatible with the forecasting horizon. In STLF, the primary inputs may include weather variables, calendar variables, and the load of the preceding hours \cite{drezga1998input,xie2017variable}. In \cite{hong2011naive,wang2016electric}, the authors consider an additional variable to address
locally increasing (or decreasing) load and the fact that electricity demand is affected by the temperatures of the preceding hours, respectively. Therefore, the quantity and quality of input data be important for an accurate estimation.
Smart grid and renewable energy resources bring opportunities for improved forecasts by providing sensing and data acquisition capabilities with high resolution \cite{kezunovic2020big}. This work discusses big data applications and associated implementation issues in load forecasting. \par
With these new sources of data, the input data may face challenges, including measurement device error and cyberattacks (e.g., false data injection) that cause operators to make improper decisions. The vulnerability analysis for input data in load forecasting is analyzed in \cite{chen2019exploiting}. It is a vital task to check the quality of input data to guarantee accuracy. Statistical analysis and information theory \cite{chandola2009anomaly}, machine learning (ML)-based method \cite{cui2019machine,esmalifalak2014detecting} are some approaches for anomaly detection schemes. \par 
As the attackers also benefit from AI and ML methods, attackers may deploy sophisticated techniques to bypass the anomaly detection, such as, targeting state estimation and electricity market application \cite{liu2011false,xie2017variable, zhang2021cyber}. Although detection scheme imposes difficulty for the attacker, this may be overcome by, for example, reinforcing the threat model \cite{chakraborty2018adversarial}. To address this issue, physical-based constraints can provide a stronger obstacle for attacker. Li et al. propose adding inherent constraints in power system  state estimation (SE) \cite{li2021conaml}. This idea makes a further constrained problem that needs to be solved by attackers while satisfying the physical systems' inherent constraints. In this paper, we seek to deploy a similar concept that supports building a defense mechanism for STLF against adversarial attack. \par
Load forecasts can be represented as a time series, which offers various tools to identify characteristics. In this regards, similarity measures can facilitate us to identify physical-based constraints. The paper's main contribution is that this is a first  attempt to define a protective mechanism in STLF associated with these inherent constraints. \par 
The paper is organized as follows: The STLF formulation and modeling are presented in Section \ref{section:1}. Section \ref{section:2} presents the problem statement in STLF capturing the constraints. Exploring constraints using similarity measures is explained in Section \ref{section:2}. Then, simulation results and observations are discussed in Section \ref{section:4}. Section \ref{section:5} provides some concluding remarks. 
\section{Short term load forecasting modeling}\label{section:1} 
In STLF, the objective is to estimate the load for a short time horizon, typically, ranging from the next hour to one week. The main intention of STLF is an hourly prediction of aggregated load. In addition, the STLF may focus on forecasting the daily peak system load, the value of system load at a specific time of day, hourly or half-hourly of system energy, and the daily and weekly system energy \cite{gross1987short}. In this regard, the STLF is concerned with scheduling purposes for the most economic commitment of generation sources consistent with reliability requirements, operational constraints. Furthermore, the prediction of load may be used for the assessment power system security for the off-line network analysis under which the part of systems may face stress. \par
The system load is the sum of all the individual demands in all nodes of power system. Many factors impact the STLF model of aggregate load. The factors are mainly categorized into weather, calendar day, economic, social events and other random factors. So, the dependent variable that needs to be estimated is aggregate load, and other variables relating to effective factor categories are considered as independent variables (explanatory). A practical report indicating independent variables for different load-serving entities (LSEs) is found in \cite{carvallo2016load}. \par
The general formulation of STLF is to define a forecasting function \(f\) as 
$y_{t}= f(\bar{X}_{t})$ where $y_{t}$, and $\bar{X}_{t}$ are the load and the independent variables at time \(t\). 
There are many ways to find \(f\), which are commonly divided into statistical and AI methods. The former is statistical analysis such as MLR, ARMA, and exponential smoothing models. An AI approach may use an ensemble of machine learning algorithms to forecasts, such as, ANN, fuzzy regression models, SVM and so on. A high-level comparison of load forecasting techniques is presented in \cite{hong2016probabilistic}. In this paper, the MLR is chosen to apply the proposed methods, although it should be generally applicable to other techniques. MLR is a broader class of regressions that encompasses linear and nonlinear regressions with multiple independent variables $\Vec{X}$ and one dependent variable $Y$. The general linear regression model can be defined as
\begin{align}\label{mlr}
Y= \beta_{0}+ \beta_{1}  X_{1} + \beta_{2}  X_{2} +\beta_{3}  X_{3} + \dots +  \beta_{n}  X_{n} +\epsilon  
\end{align}
where $\beta_{0},\beta_{1}, \dots$ are the parameters and $\epsilon$ is normal random error. Note that second or higher-ordered terms of a independent variables is equivalent to the first ordered one and it is a special case of Eq. \ref{mlr} \cite{hong2011naive}. \par
\subsection{Vulnerability of independent variables}
Some independent variables cannot be easily corrupted as their values are widely known. This includes calendar variables (month of the year, day of the week, hour of the day, etc.). Economic factors are not explicitly expressed in STLF models due to the longer time scale. Among other variables, weather factors significantly in the load pattern because many loads are sensitive to temperatures, such as, heating and cooling. \par
The temperature data is the most used features in many STLF models to capture the load behavior. In this case, the quality of temperature is crucial to STLF accuracy. The temperature used in the STLF model is an estimated value where is obtained from commercial weather services or external services/APIs. This also provides a vulnerable area for data disruptions and false attack injections. Sobhani et al. suggest temperature anomaly detection with the help of local load information collected by power companies \cite{sobhani2020temperature}. Although the detection scheme can be helpful to find the anomaly in temperature, an attacker who also benefits from AI methods would be more challenging. The adversarial attack on temperature data exploited in \cite{chen2019exploiting} indicates that STLF models are quite vulnerable to malicious data in temperature from online weather services. In addition, an attacker could manipulate load forecasts in arbitrary directions and cause significant and targeted damages to system operations. \par
\section{Problem statement} \label{section:2}
Considering the temperature as a targeted variable for the attacker, we investigate constraints to act as a defense mechanism such that the attacker must satisfy these in order to have a feasible result. Being undetected from system operators is the primary task for the attacker to generate an adversarial action. Generally, the attacker's capability will be limited for a successful malicious action. That means the difference between corrupted temperature and actual temperature will be bounded \cite{chen2019exploiting}. Considering $H$ is attacker simulated forecasting model parameterized by $\theta$. The only constraint here is about bypassing the detection scheme. Adding the psychical-based constraints may cause difficulty for the attacker to find a feasible attack vector. Hence, the problem that the attacker should solve can be defined as the following optimization problem 
\vspace{-.02 in}
\begin{subequations}
\begin{align} \label{problem}
\min_{\tilde{X}} \quad &H_{\theta}(\tilde{X}) \\
 s.t. \quad  &{\left \|X-\tilde{{X}} \right\|_{p} \leq \epsilon} \label{detect} \\
 &{ g(x) \leq 0} \label{const}
 \end{align}
\end{subequations}
where $\tilde{X} , X $ are the injected and actual temperature data, $\epsilon$ is a threshold value, and \(p\) shows different norm according to detection algorithms. Here we are adding (\ref{const}) to (\ref{problem}) to represent the psychical-based constraints. The minimization problem here points out attacker target as decreasing the load forecast values. The constraint $g(x)$ can be imposed either as an equality or inequality related to the physical system and topology. The right side of (\ref{const}) may be defined by an upper bound to deal with various scenarios. The way to find $g(x)$ is discussed in the next section.
\section{Exploring Constraints for STLF}\label{section:3}
Power systems follow some physical and topological constraints. For instance, output of power metering devices should ideally capture Kirchhoff’s laws. This action may present a built-in defense mechanism. Li et al. study potential vulnerabilities of ML applied in power systems by proposing constraints that adversarial examples must satisfy the intrinsic constraints of physical systems \cite{li2021conaml}. In power system SE the physical constraints have already been encoded in the mathematical model; however Liu et al.  propose an approach that guarantees attacker to pass those constraints in linear case \cite{liu2011false}. \par
The indication of such constraints related to load forecasting applications has not been investigated in the literature. Since STLF mainly applies to aggregated load models, the dynamic of loads that follow the physics law is not easily observed. To build a set of psychical-based constraints, the spatial distribution of zones is considered. For example, ERCOT load map can be separated based on eight different weather zones. Each zone deploys its weather station to STFL, which may also represent an aggregate. \par
Generally speaking, we investigate the relation between the different zones to derive constraints. In this framework, we are looking for an index representing forecasted variations between zones. When an attacker intents to inject false data into some weather stations (but not all), the malicious action may be identified by these indices. Here, similarity metrics applied in clustering time series are inspected to find a relation between zones. In fact, the goal is to explore the physical-based constraints between zones $g(x)$ that may be obtained by these similarity measures.
\begin{definition}
  The \textit{similarity measure $D(X, Y)$} between time series $X=\{x_{0}, x_{1}, ..., x_{N\text{-}1}\}$ and $Y=\{y_{0}, y_{1}, ..., y_{N\text{-}1}\}$ is a function taking two time series as inputs and returning the \textit{distance \(d\)} between these series \cite{esling2012time}.
\end{definition} \par
Two common approaches for similarity measure are considered to deploy as shape-based and feature-based methods.
\vspace{-.1 in}
\subsection{Shape-based approach}
Here, distances are based on directly comparing the raw values and the shape of the series in different manners. That is, distances compare the overall shape of the series. Due to the simplicity, the Euclidean distance $d_{EUC}$ and other $L_{p}$ norms are the most widely used distance measures for time series clustering. Euclidean distance is invariant when dealing with changes in the order that time features are presented. The Euclidean distance is defined as 
\begin{align}
d_{EUC}=\sqrt{\sum_{i=0}^{N-1}\left(x_{i}-y_{i}\right)^{2}}
\end{align} 
\vspace{-.05 in}
\subsection{Feature-based approach}
The feature-Based method focuses on extracting a set of features from the data presented in time series and calculating the similarity between the features instead of using the raw values of the series.
\subsubsection{Correlation-based distance}
 One index is to use the Pearson's correlation factor for two time series. The time series are similar if they are highly correlated, even though the observed values may be far apart in terms of Euclidean distance. Based on Pearson's correlation, Golay et al. suggest two correlation-based distances for time series $X$ and $Y$ as follows \cite{golay1998new} 
\begin{align} \label{d_ecu}
d_{COR\cdot 1}\left({X}, {Y}\right)&=\sqrt{2\left(1-C O R\left({X}, {Y}\right)\right)} \\
d_{C O R .2}\left({X}, {Y}\right)&=\sqrt{\left( \frac{1-\operatorname{COR}\left({X}, {Y}\right)}{1+\operatorname{COR}\left({X}, {Y}\right)}\right)^{\beta} } 
\end{align}
where $\beta>0$ is a distance decreasing parameter, and $COR$ is the correlation between $X$ and $Y$.
It can be observed that a strong positive correlation results in small-scale correlation-based distances, which implies high similarity.
\subsubsection{Autocorrelation-based distance}
The other feature-based method is based on estimated autocorrelation functions. The autocorrelation function is the normalized autocovariance function indicating the estimated auto correlation vectors of two times series as $\hat{\rho}_{X_T}$ and $\hat{\rho}_{Y_T}$, the autocorrelation-based distances between $X$ and $Y$ is defined as follows \cite{galeano2001multivariate}
\begin{align} \label{d_acf}
d_{A C F}\left({X}, {Y}\right)=\sqrt{\left(\hat{{\rho}}_{X_{T}}-\hat{{\rho}}_{Y_{T}}\right)^{\top} {\Omega}\left(\hat{{\rho}}_{X_{T}}-\hat{{\rho}}_{Y_{T}}\right)}
\end{align}
where ${\Omega}$ represents weight matrix, and \(T\) is the sample size.
\subsubsection{Periodogram-based distance approach} This method computes the distance between periodograms of two time series in the frequency domain. A periodogram is an estimate of the spectral density of a signal. For a stationary time series process with finite length, the periodogram at the Fourier frequency $\omega_{j}= (2\pi j)/N$ for $j= 1,2, \dots,[N/2]$ (where $[N/2]$ is the largest integer less or equal to $n/2$) is defined as 
\begin{align}
    \boldsymbol{P}_{x}(\omega) = \frac{1}{N} |\sum_{t=1}^{N}{x_{t} e^{-it\omega_{j}}}|^{2}
\end{align}
The scaled periodogram distance can be defined as 
\begin{align}
d_{\mathrm{P}}(X, Y)=\sqrt{\sum_{j=1}^{[N / 2]}\left[\boldsymbol{P}_{x}\left(w_{j}\right)-\boldsymbol{P}_{y}\left(w_{j}\right)\right]^{2}}
\end{align}
The normalized version gives more weight to the shape of the curve, while the non-normalized contributes to the scale. Measures based on the autocorrelations and measures based on the periodogram are related. However, this application could provide different useful estimation results.
\subsubsection{Symbolic representation}
Transformation of time series into sequences of discretized symbols can be efficiently processed to extract information about the underlying processes. The Symbolic Aggregate approXimation (SAX) method proposed in \cite{lin2003symbolic} allows to investigate the dissimilarity based on the symbolic representation of time series to find the true dissimilarity between the original time series. More details about the procedure are addressed in \cite{lin2003symbolic}.

\section{Illustration and Simulation }\label{section:4}
 \subsection{Dataset}
\begin{figure}[h!]
    \centering
    \includegraphics[scale=0.43]{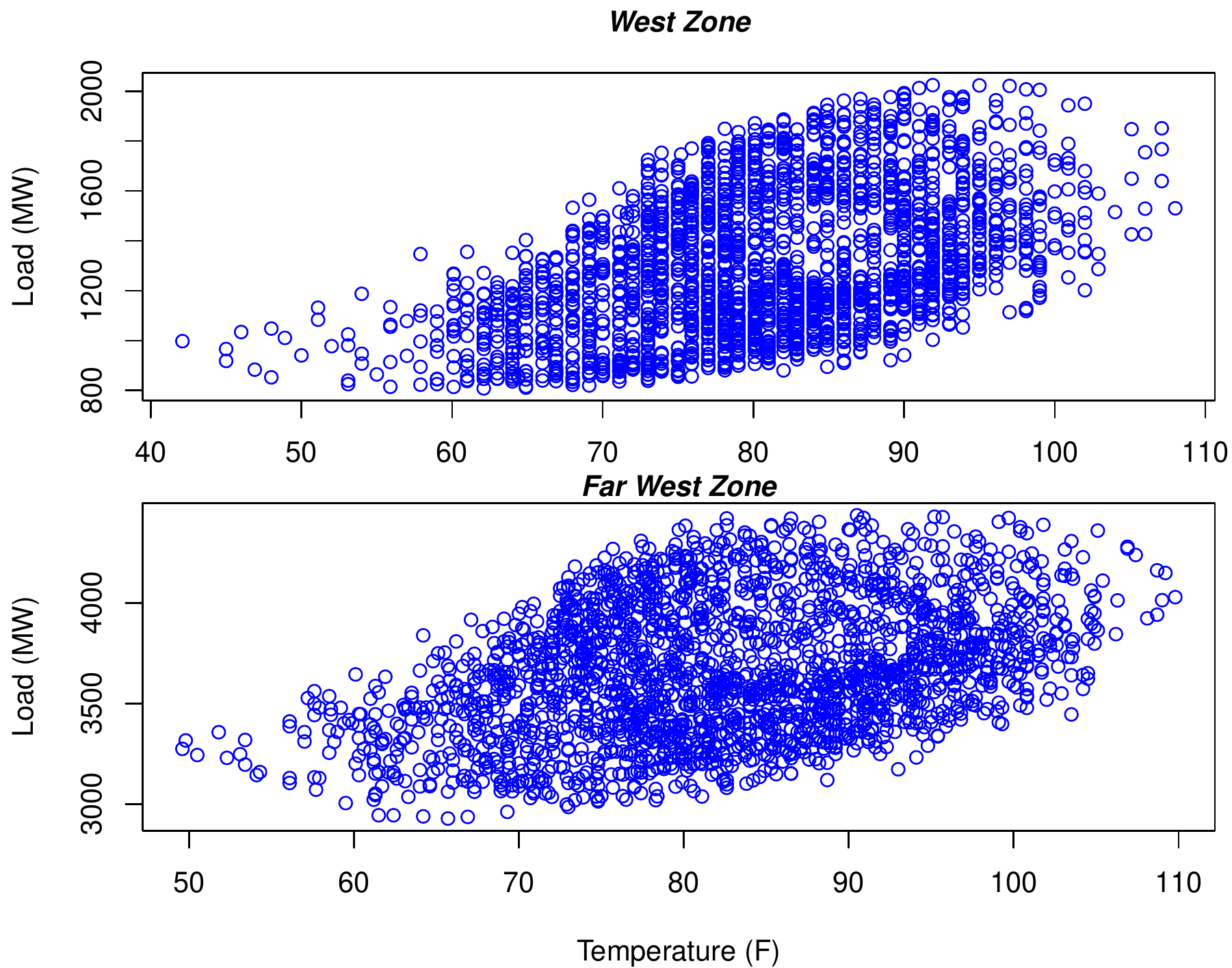}
    \vspace{-.1 in}
    \caption{Load vs. temperature}
    \label{fig:summer_temp_load}
\end{figure}

The ERCOT area can be categorized based on eight weather stations, where two West and Far West areas are selected to apply the proposed methods. The load data is for summer 2020, and it is accessible from the ERCOT website. This data set includes the overall and individual area load. As the temperature is an essential variable in STLF modeling, the temperature data was collected from Automated Surface Observing Systems (ASOSs) corresponding to the West and Far West weather stations. After data cleaning steps such as handling missing values, removing unwanted observations, particularly for the temperature data set, the recorded load verse temperature is shown in Fig. \ref{fig:summer_temp_load}. As it is clear, high temperature causes increasing use of demand-side devices such as air conditioners. 
\vspace{-0.15 in}
\vspace{-.2 in}
\subsection{Correlation Analysis}
\begin{figure}[h!]
    \centering
    \includegraphics[scale=.45]{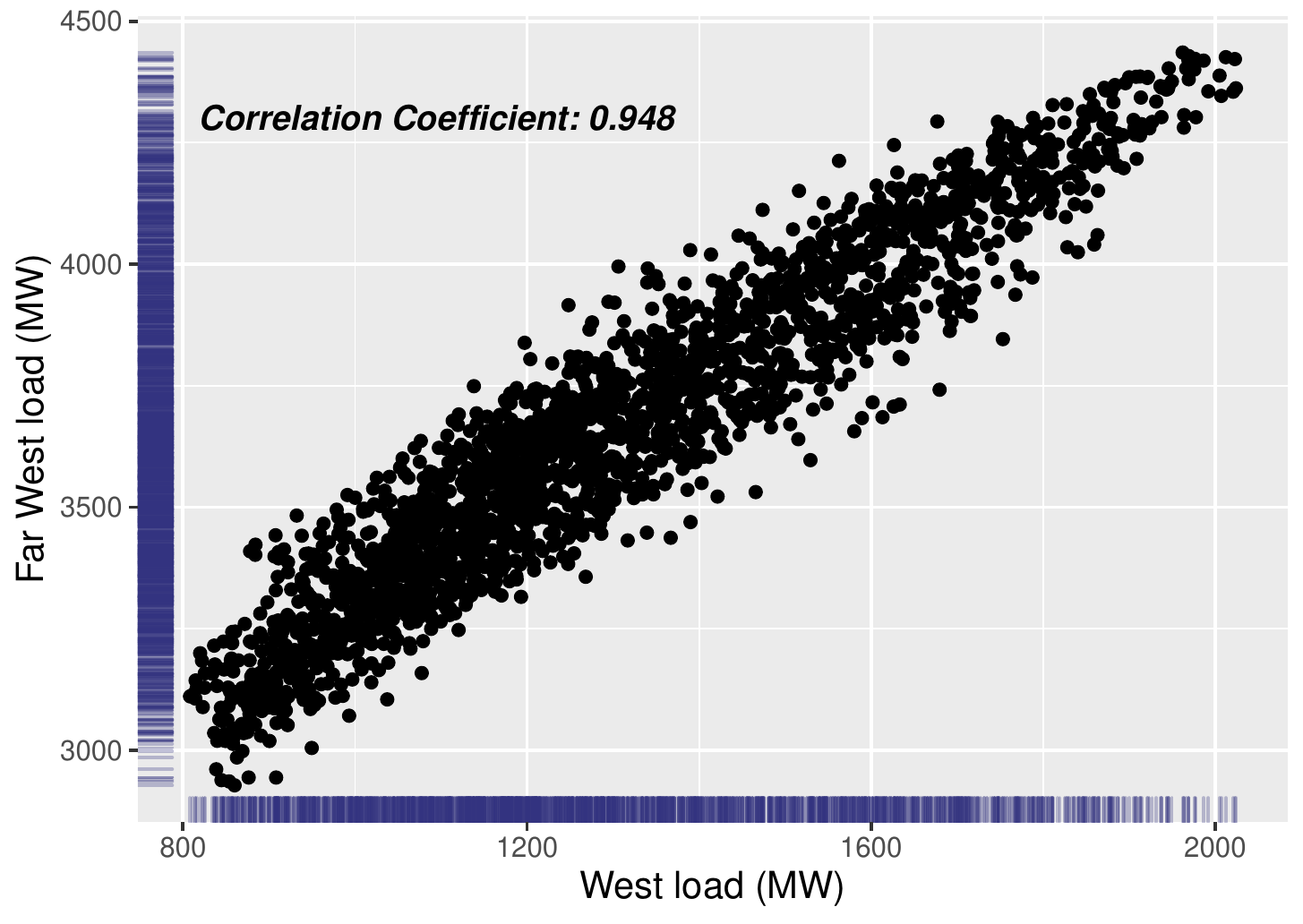}
    \vspace{-.1 in}
    \caption{West and Far West zone load correlation }
    \label{fig:load corr}
\end{figure}

To explore possible relations between loads in the west and far west in ERCOT, first, the correlation analysis between load data is calculated, and results are shown in Fig. \ref{fig:load corr}. Although the two zones are separated geographically, there is a strong correlation between these loads. The reason could be the similarity of weather patterns in each zone that demonstrates hidden physical associations.
\subsection{STLF modeling and results}
Two MLR models are applied for STLF modeling since the impact of temperature data as an independent variable may be seen explicitly in the formulation. Note that other techniques, either statistical or AI methods, are applicable in this case. Two models including interaction effect of independent variables based on  \cite{hong2011naive} are assumed as 
\begin{align}
f_{1}&= \beta_{0}+ \beta_{1} \times T + \beta_{1} \times H +\beta_{2}\times D +\beta_{3} \times LL_{1w}  \\&+  \beta_{4}  \times LL_{2w} \nonumber  \\
f_{2}&= \beta_{0}+ \beta_{1}\times D \times H + \beta_{2} \times M \times T +\beta_{3} \times M \times T^{2}  \nonumber \\ &+\beta_{3} \times M \times T^{3}  + \beta_{4} \times H \times T + \beta_{5}\times H \times T^{2} \nonumber \\ &+\beta_{6} \times H \times T^{3}
\end{align}
where $f_{1}$, and $f_{2}$ are the foretasted load models. The model parameters $\Vec{\beta}$ are obtained based on the least-squares (LS) method. The variables $D$, $M$, and $H$ denote the day of the week, excluding the weekend, month of the year, and hour of the day, respectively. $T$ is the temperature value, and $LL_{1w}$ and $LL_{2w}$ express one and two-week lagged load data, respectively. \par
\begin{figure}[t]
    \vspace{-.1 in}
    \centering
    \includegraphics[width=\columnwidth,height=4.2 in]{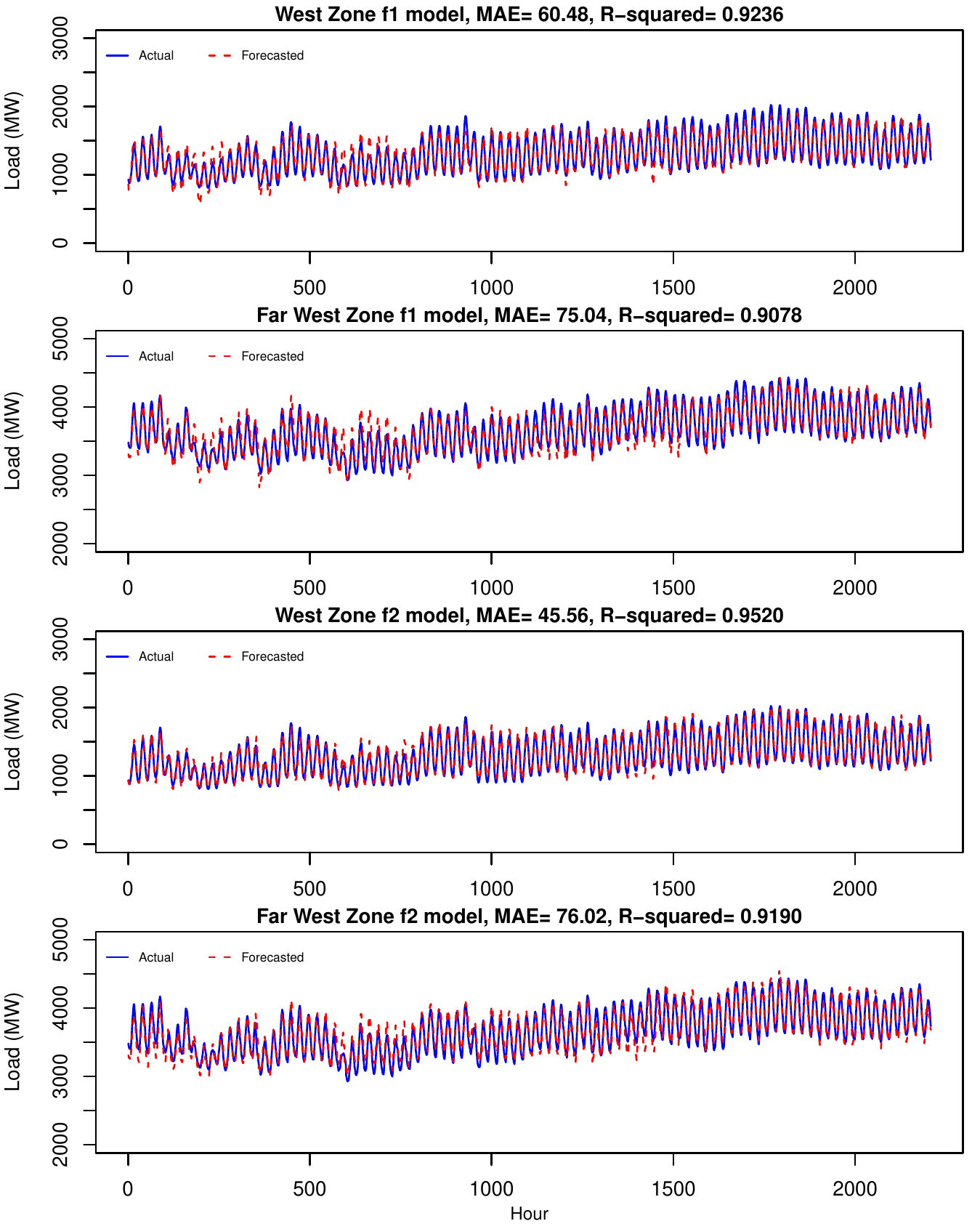}
    \vspace{-.3 in}
    \caption{Models performance: Mean absolute error (MAE) and adjusted R-squared values}
    \label{fig:model_result}
\end{figure}

Both models consist of continuous variables and categorical variables. The categorical variables need to be coded into a series of variables in order to be used in regression modeling. So, variables $D$, $M$, and $H$ are encoded as a constant factor in the R programming language. The data set is randomly divided into 70\% and 30\% for train and test set, respectively. The models result are shown in Fig. \ref{fig:model_result}.
The outcomes indicate that either $f_{1}$ or $f_{2}$ model could efficiently represent the load for both zones. A Gaussian noise with zero mean and unit variance is injected into West zone temperature data to mimic malicious actions. Then, similarity measure methods for the forecasted load are applied.\par
Table \ref{tab4} indicates the variation of similarity measure indices in the presence of the attack. Note that comparing the results between indices is misleading since each measure carries a different scale. The similarity measures change when there is false injection data. Thus, this may be a suitable candidate for constraints in STLF. Among the methods, the variant in $d_{SAX}$ is the largest. The reason would be distance measure defined on the symbolic approach creates lower bound corresponding distance measures defined on the original time series. The variation in measures is not a constant value and will vary under different scenarios, such as different Gaussian signals. That is finding an explicit formula or the upper bound capturing different scenarios is an essential. In this case, to inject data with a destructive impact on the STLF, attackers must build an attack vector to bypass the measures satisfying the STLF similarity measures . 
\begin{table}[t]
\centering
\vspace{-.2 in}
\caption{Similarity measures results}
\label{tab4}
\small
\resizebox{\columnwidth}{!} 
{
\begin{tabular}{@{}cccccc@{}}
\toprule
\multirow{2}{*}{Method} &  & \multicolumn{2}{c}{No False Injection} & \multicolumn{2}{c}{False Injection} \\ \cmidrule(l){2-6} 
 & No Model & $f_{1}$  & $f_{2}$  & $f_{1}$  & $f_{2}$  \\ \midrule
$d_{ECU}$ & 110644.5 & 110587.1 & 110675.4 & 110652.9 & 110602.3 \\
$d_{COR}$ & 0.3204611 & 0.3183464 & 0.2687631 & 0.326226 & 0.2826415 \\
$d_{ACF}$ & 1.247954 & 1.146977 & 1.015968 & 1.149789 & 1.040784 \\
$d_{p}$ & 0.1336534 & 0.1309215 & 0.1109201 & 0.1273667 & 0.1091463 \\
$d_{SAX}$ & 2.004495 & 1.735943 & 1.417392 & 2.454994 & 1.002247 \\ \bottomrule
\end{tabular}
}
\end{table}
\section{Conclusion}\label{section:5}
STLF plays an important role in power system operation. Collecting temperature, as a critical input from online services, increases the possibility of a malicious action. This paper examines some physical constraints to address deficiency of detection schemes. A new formulation of STLF based on different spatial coordination is proposed for imposing these constraints. Similarity measure techniques are applied as constraints in the proposed approach. Simulation results indicate that the applied measures can be applicable for encoding constraints when temperature data is corrupted. Distance defined in symbolic space identifies variation that can represent a strict constraint following the proposed STLF topology. Since the exploited measures vary based on scenarios, an attempt to define the explicit equation representing the constrains or to realize the upper bound will be investigated for the feature work. As part of our ongoing work, we are investigating incorporating such constraints directly into the learning algorithms for load forecasting to mitigate cyber attacks. 

\bibliographystyle{IEEEtran}
\bibliography{Ref}  
\end{document}